\documentclass[aps,prl,preprint,nopacs,superscriptaddress]{revtex4}

\usepackage{graphicx}
\usepackage{verbatim}
\usepackage{mathrsfs}
\usepackage{amsmath,amsfonts,amssymb}
\usepackage{graphicx}

\def\3{2.8in}    
\def\2{2.5in}
\def\4{3.0in}

\def \beq {\begin{equation}}
\def \eeq {\end{equation}}
\begin{document}

\title{Observation of the Adler-Bell-Jackiw chiral anomaly in a Weyl semimetal}

\author{Chenglong Zhang\footnote{These authors contributed equally to this work.}}\affiliation{International Center for Quantum Materials, School of Physics, Peking University, China}
\author{Su-Yang Xu$^*$}\affiliation {Joseph Henry Laboratory, Department of Physics, Princeton University, Princeton, New Jersey 08544, USA}
\affiliation{Princeton Center for Complex Materials, Princeton Institute for the Science and Technology of Materials, Princeton University, Princeton, New Jersey 08544, USA}
\author{Ilya Belopolski$^*$}\affiliation {Joseph Henry Laboratory, Department of Physics, Princeton University, Princeton, New Jersey 08544, USA}
\affiliation{Princeton Center for Complex Materials, Princeton Institute for the Science and Technology of Materials, Princeton University, Princeton, New Jersey 08544, USA}
\author{Zhujun Yuan$^*$}\affiliation{International Center for Quantum Materials, School of Physics, Peking University, China}
\author{Ziquan Lin}\affiliation {Wuhan National High Magnetic Field Center, Huazhong University of Science and Technology, Wuhan 430074, China}
\author{Bingbing Tong}\affiliation{International Center for Quantum Materials, School of Physics, Peking University, China}
\author{Nasser Alidoust}\affiliation {Joseph Henry Laboratory, Department of Physics, Princeton University, Princeton, New Jersey 08544, USA}
\author{Chi-Cheng Lee}
\affiliation{Centre for Advanced 2D Materials and Graphene Research Centre National University of Singapore, 6 Science Drive 2, Singapore 117546}
\affiliation{Department of Physics, National University of Singapore, 2 Science Drive 3, Singapore 117542}
\author{Shin-Ming Huang}
\affiliation{Centre for Advanced 2D Materials and Graphene Research Centre National University of Singapore, 6 Science Drive 2, Singapore 117546}
\affiliation{Department of Physics, National University of Singapore, 2 Science Drive 3, Singapore 117542}
\author{Hsin Lin}
\affiliation{Centre for Advanced 2D Materials and Graphene Research Centre National University of Singapore, 6 Science Drive 2, Singapore 117546}
\affiliation{Department of Physics, National University of Singapore, 2 Science Drive 3, Singapore 117542}
\author{Madhab Neupane}\affiliation {Joseph Henry Laboratory, Department of Physics, Princeton University, Princeton, New Jersey 08544, USA}
\affiliation {Condensed Matter and Magnet Science Group, Los Alamos National Laboratory, Los Alamos, New Mexico 87545, USA}
\author{Daniel S. Sanchez}\affiliation {Joseph Henry Laboratory, Department of Physics, Princeton University, Princeton, New Jersey 08544, USA}
\author{Hao Zheng}\affiliation {Joseph Henry Laboratory, Department of Physics, Princeton University, Princeton, New Jersey 08544, USA}
\author{Guang Bian}\affiliation {Joseph Henry Laboratory, Department of Physics, Princeton University, Princeton, New Jersey 08544, USA}
\author{Junfeng Wang}\affiliation {Wuhan National High Magnetic Field Center, Huazhong University of Science and Technology, Wuhan 430074, China}
\author{Chi Zhang}
\affiliation{International Center for Quantum Materials, School of Physics, Peking University, China}\affiliation{Collaborative Innovation Center of Quantum Matter, Beijing,100871, China}
\author{Titus Neupert}\affiliation {Joseph Henry Laboratory, Department of Physics, Princeton University, Princeton, New Jersey 08544, USA}
\affiliation {Princeton Center for Theoretical  Science, Princeton University, Princeton, New Jersey 08544, USA}
\author{M. Zahid Hasan\footnote{Corresponding authors (emails): mzhasan@Princeton.edu and gwljiashuang@pku.edu.cn}}\affiliation {Joseph Henry Laboratory, Department of Physics, Princeton University, Princeton, New Jersey 08544, USA}
\affiliation{Princeton Center for Complex Materials, Princeton Institute for the Science and Technology of Materials, Princeton University, Princeton, New Jersey 08544, USA}
\author{Shuang Jia$^\dag$}
\affiliation{International Center for Quantum Materials, School of Physics, Peking University, China}\affiliation{Collaborative Innovation Center of Quantum Matter, Beijing,100871, China}


\date{\today}

\begin{abstract}
\textbf{Relativistic fermions can be of three important varieties: Dirac, Majorana and Weyl. A Weyl semimetal is a novel quantum solid whose electrons behave like emergent massless fermions with definite handedness or chirality known as the Weyl fermion which is distinct from Dirac or Majorana fermions. Among all the unusual phenomena that are enabled by such quantum solids, the chiral anomaly is the most striking one because it apparently violates the conservation of charge for the particle number with a given chirality. Here, for the first time, we report experimental studies of the first Weyl semimetal TaAs which reveals the chiral anomaly in its magnetotransport. Unlike most metals that become more insulating or resistive under an external magnetic field, we observe that our high mobility TaAs samples, quite remarkably, become more conductive as a magnetic field is applied along the direction of the current for certain ranges of the field and its magnetoconductance disperses quadratically which is nearly independent of temperatures below 20 K, but depends strongly on the relative angles between the electric and magnetic fields. These systematic results, corroborated by additional observations and our theoretical calculations here, collectively suggest the existence of chiral anomaly or the non-conservation of charge for a given chiral channel in TaAs analogous to anomalies in quantum field theory. Our results not only provide the first transport signature for the Weyl fermions in nature, but also paves the way for utilizing chiral fermions in applications such as the `valleytronic' devices which is a long-sought platform for the next generation technology frontier.
}
\end{abstract}
\maketitle

\begin{figure*}
\centering
\includegraphics[width=17cm]{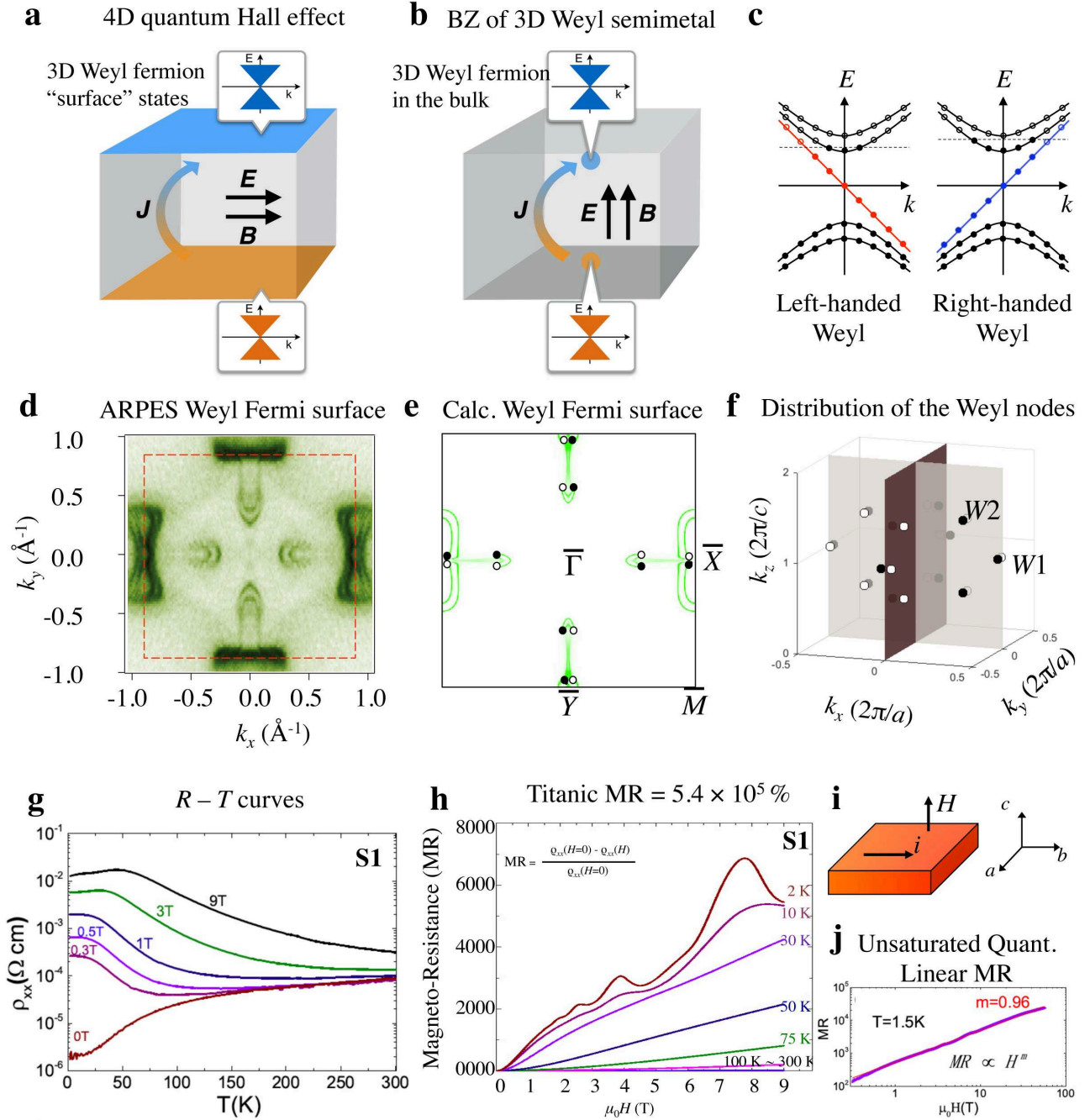}
\caption{\label{Fig1}\textbf{The Weyl semimetal state and titanic magneto-resistance in TaAs.} \textbf{a,} Schematics of the separation the Weyl fermions with opposite chiralities in a 4D quantum Hall effect. Here, the 3D Weyl fermions surface states  with  opposite chiralities reside on the opposite surfaces. \textbf{b,} Schematics of the separation of the pairs of Weyl fermions in a Weyl semimetal with opposite chiralities in momentum space, which is a direct consequence of its nontrivial topological nature. \textbf{c,} Schematics of pumping electrons from one Weyl cone to another one with the opposite chirality, is separated in momentum space, in the presence of parallel magnetic and electric fields. The red}
\end{figure*}

The laws of physics rest crucially on symmetries and the associated conservation laws. Over the last century, physicists have repeatedly observed violations of conservation laws in particle physics, each time leading to revolutions in our understanding of the basic laws of nature. One of the most striking phenomena of this type is the breaking of a conservation law of classical physics by quantum mechanical effects, a so-called anomaly \cite{Anomaly}. Perhaps the most well-known example is the so-called chiral anomaly associated with Weyl fermions \cite{Nielsen1983, ABJ1, ABJ2, Volovik2003}. A Weyl fermion is a massless fermion that carries a definite chiral charge. Due to the chiral anomaly, the chiral charge of Weyl fermions is not conserved by quantum fluctuations. Realizing a condensed matter analog of the chiral anomaly not only adds another remarkable example (analogous to Dirac and Majorana fermion) to the correspondence between particle and condensed-matter physics, but also may lead to new phenomena and devices in solid state materials that are uniquely enabled by such an exotic quantum anomaly. In order to realize the chiral anomaly, that is to say, the violation of the conservation of chiral current, one needs to have a solid state system whose low energy excitations consists of Weyl fermions and to separate the Weyl fermions of opposite chirality. A straightforward route is to separate them in real space. This was proposed in a 4D quantum Hall effect \cite{4DQHE}, where the ``surface'' states of this 4D objects are Weyl fermions that have the opposite chiralities on the opposite surfaces (Fig.~\ref{Fig1}\textbf{a}). In this case, the Hall current manifests the chiral anomaly because it transfers charges from one surface to the other, apparently thus violating the charge conservation on either surface. However, due to the absence of an experimental access to a 4D space, this is impossible to realize.

On the other hand, recent theoretical advances in topological physics has predicted that Weyl fermions can arise in the bulk of certain novel semimetals with nontrivial topology \cite{Balents_viewpoint, Wan2011, Murakami2007, TI_book_2014, Ashvin_Review, HasanWeyl, XDai}. A Weyl semimetal is a bulk crystal whose low energy excitations satisfy the Weyl equation. Therefore, the conduction and valence bands touch at discrete points, the Weyl nodes, with a linear dispersion relation in all three momentum space directions moving away from the Weyl node. Now, although in this case one cannot separate the Weyl fermions with opposite chiralities in real space, however the nontrivial topological nature of a Weyl semimetal guarantees that pairs of Weyl fermions with the opposite chiralities are separated in momentum space (Fig.~\ref{Fig1}\textbf{b}). Therefore, one can imagine that electrons can be pumped from one Weyl cone to another one with the opposite chirality that is separated in momentum space, in the presence of parallel magnetic and electric fields (Figs.~\ref{Fig1}\textbf{b} and~\ref{Fig1}\textbf{c}), which violates the conservation of the particle number with a given chirality \cite{Nielsen1983, Chiral_Burkov, Qi_review}. This gives rise to a novel analog of the chiral anomaly in a condensed matter system. Apart from this elegant analogy and correspondence between condensed matter and high energy physics, which by itself is of great interest, the chiral anomaly also serves as a crucial transport signature for Weyl fermions in a Weyl semimetal phase. Furthermore, theories have recently predicted its potential applications in `valleytronic' devices \cite{Nonlocal}. Despite these, the chiral anomaly in condensed matter bulk solids has not been observed because a material realization of the Weyl semimetal phase remained elusive for many many years. Very recently, photoemission (ARPES) experiments have provided strong evidence for the first Weyl semimetal realization in an inversion breaking, stoichiometric solid, TaAs \cite{Huang2015, Weng2015, TaAs Hasan, TaAs Ding}. Here, we measure the condensed matter chiral anomaly in the Weyl semimetal TaAs by directly observing its key signature in the form of a quadratic positive large magnetoresistance. We further show that such a positive magnetoconductance has a very weak temperature dependence below 20~K, but depends strongly on the relative angle between the electrical and magnetic fields.  These systematic results, corroborated by our other observations and theoretical calculations, collectively demonstrate the chiral anomaly in the Weyl semimetal compound TaAs.

We start by studying the surface electronic structure of the TaAs samples that are used in our electrical transport experiments by angle-resolved photoemission spectroscopy (ARPES), because a Weyl semimetal is a gapless topological state of matter described by a unique bulk-boundary correspondence and because ARPES is the most natural and direct probe of surface state band structure, Fermi surface, and topological properties. Figure~\ref{Fig1}\textbf{d} shows our ARPES measured Fermi surface of the (001) surface. Specifically, in the vicinity of each $\bar{X}$ point, we observe a bow-tie shaped surface state; Near each midpoint between $\bar\Gamma$ and $\bar{X}$ or $\bar\Gamma$ and $\bar{Y}$, we observe a pair of open curves, Fermi arcs, that are terminated into two points, which are the Weyl nodes. These observations are in qualitatively consistent with our first-principles calculation shown in Fig.~\ref{Fig1}\textbf{e}. The observation of Fermi arc surface states and the agreement with the theoretical calculation provide strong evidence showing that the TaAs samples used in our transport experiments are indeed Weyl semimetals. Since transport mostly probes the bulk electronic states, we need to understand the electronic structure of the bulk Weyl cones. In Fig.~\ref{Fig1}\textbf{f} we show distribution of the Weyl nodes throughout the first bulk Brillouin zone (BZ). There are in total 24 Weyl nodes and the sign of their chiral charges are color coded in black and white. Moreover, these 24 Weyl nodes can be characterized into two groups. Namely, there are 8 Weyl nodes located on the $k_z=\pi$ plane that have the same energy, and we note them as W1. The other 16 Weyl nodes, which are away from the $k_z=\pi$ plane, are noted as W2.

After checking the electronic ground state of our TaAs samples, we study their electrical transport properties. Fig.~\ref{Fig1}\textbf{g} shows the longitudinal resistivity as a function of temperature at different external magnetic fields. The current is applied along the in-plane crystallographic direction ($a$ and $b$ are equivalent because of the tetragonal lattice) whereas the magnetic field is out-of-plane. The temperature dependent resistivity of TaAs at the zero magnetic field (Fig.~\ref{Fig1}\textbf{g}) shows a metallic profile. When a low magnetic field (0.3~T) is applied, the resistivity changes to an insulating profile. Similar behavior has been observed in bismuth and graphite, where the low-field effect was attributed to a magnetic-field-induced excitonic insulator transition of Dirac fermions \cite{MITchinese}. In Fig.~\ref{Fig1}\textbf{h}, we show the magnetoresistance [$\mathrm{MR}\equiv(\rho_{H=0}-\rho_{H})/\rho_{H=0}$] as a function of the external magnetic field at different temperatures. At temperatures below 10~K, the $\mathrm{MR}$ of TaAs shows extremely strong Shubnikov-de Haas (SdH) oscillations. Remarkably, the $\mathrm{MR}$ reaches 5400 at 10~K in 9~T, which is three times larger than that in the recently reported compound WTe$_2$ at 2~K in 9~T (see, Nature (2014)) \cite{ali_WTe2_2014}. Moreover, unlike in other semimetals where the MR has a parabolic dependence on the magnetic field, the titanic $\mathrm{MR}$ of TaAs disperses linearly as a function of the magnetic field and is not saturated at the highest applied field of 56~T (Fig.~\ref{Fig1}\textbf{j}).

In order to understand the electronic states that contribute to our transport signals, we study the SdH oscillation. Figure~\ref{Fig2}\textbf{a} shows the Hall resistance $\rho_{yx}$ at different temperatures. It can be seen that the $\rho_{yx}$ shows a positive magnetic field response at temperatures above 100~K, which reveals that the majority carriers are hole-like at these higher temperatures. By contrast, the field response becomes clearly negative below 75~K, which shows the dominance of electron-like carriers in the Hall measurements at low temperatures. In order to obtain the carrier density $n$ and the mobility for the electron and hole carriers, we fit the Hall conductivity tensor using a two band model (see Supplementary Information (SI) for details). As shown in Fig.~\ref{Fig2}\textbf{b}, the carrier density for both carriers are on the order of $10^{17}$~cm$^{-3}$, which is consistent with the semimetallic nature of TaAs. At temperatures above 100~K, the transport data can be fitted by considering only the hole-like band. However, as one decreases the temperature below 100~K, the electron band sets in and its mobility increases dramatically. We note that the mobility of the electron band is as high as $\mu_{\mathrm{e}}=5\times10^5\ \mathrm{cm}^2/\mathrm{V}\cdot\mathrm{s}$ at $T=2$~K, which is comparable to that in Cd$_3$As$_2$ \cite{liang2014ultrahigh}.

In order to obtain the critically important electronic band structure parameters, we analyze the SdH quantum oscillation data at $T=2$~K. We note that since at $T=2$~K the SdH oscillation is dominated by the electron carriers, the obtained band parameters will correspond to the electron-like band. We use the following expression to analyze the SdH oscillation data, $\rho_{xx}$ at $T=2$~K, for a 3D system, $\rho_{xx}={\rho_0}[1+A(B,T)\cos2\pi(F/B+\gamma)]$ \cite{murakawa2013detection}, where $\rho_0$ is the non-oscillatory part of the resistivity, $A(B,T)$ is the amplitude of the SdH oscillations, $B$ is the magnetic field, $\gamma$ is the Onsager phase, and $F=\frac{\hbar}{2\pi{e}}A_{\mathrm{F}}$ is the frequency of the oscillations. Here, $A_{\mathrm{F}}$ is the extremal cross-sectional area of the Fermi surface (FS) associated with the Landau level index $\nu$, $e$ is the electron charge, and $\hbar$ is the Planck's constant. We obtain a Fermi surface area of$A_{\mathrm{F}}=7.07\times10^{-4}\ \mathrm{\AA}^{-2}$ and a Fermi wave vector $k_{\mathrm{F}}$ is $\sqrt{A_{\mathrm{F}}/\pi} = 0.015\ \mathrm{\AA}^{-1}$. We note that since the magnetic field is parallel to the $c$ crystallographic axis, the obtained Fermi surface area corresponds to the 2D cross-section of the 3D Fermi pocket that is perpendicular to the $k_z$ direction. Then the Landau level index $\nu$ ($1/\mu_0H$) is plotted as a function of the inverse of the magnetic field strength in Fig.~\ref{Fig2}\textbf{d}, from which one can see that, for all four samples, the linear interpolation of the curve intersects with the x axis at $\nu=0$. This suggests that the electron carriers arise from a linearly dispersive band with a non-trivial Berry's phase \cite{murakawa2013detection}, which is likely the Weyl cone. In order to obtain the Fermi vector, the Fermi velocity, the energy position of the chemical potential, and other important band parameters, we apply the Lifshitz-Kosevich formula for a 3D system (see the SI). This enable us a cyclotron mass $m_{\mathrm{cyc}}$ of $0.15m_{\mathrm{e}}$. From the cyclotron mass, we obtain the Fermi wave vector $k_{\mathrm{F}}$ is $\sqrt{A_{\mathrm{F}}/\pi} = 0.015\ \mathrm{\AA}^{-1}$, and the fermi velocity $v_F$  is ${\hbar}k_{\mathrm{F}}/m_{\mathrm{cyc}} = 1.16\times10^5$~$\mathrm{m/s}$. By assuming a linear dispersion of this electron pocket, we obtain the chemical potential (relative to the energy of the Weyl node) to be $E_{\mathrm{F}}$ = $m_{\mathrm{cyc}}v_{\mathrm{F}}^2 = 11.48$~meV. We further study the anisotropy of the electron-like pocket by tilting the magnetic field away from the $c$ direction. Our data (Fig.~\ref{Fig2}\textbf{e}) shows that the Fermi surface area along the $a$ axis is about 5 times larger that of along the $c$ axis. Therefore, we find that the electron-like Fermi pocket is an ellipsoid that is elongated along the $c$ axis.

We now systematically check if the obtained band parameters make sense and if the electron-like band indeed corresponds to the Weyl cone. In order to do so, we place the chemical potential at 11.48~meV above the Weyl nodes W1 in our first-principles calculations and try to compare the calculated band parameters to those of obtained from transport. We have found an excellent agreement between calculation and transport: (1)~As shown in Figs.~\ref{Fig2}\textbf{f} and~\ref{Fig2}\textbf{g}, if the chemical potential is placed at 11.48~meV above W1, calculation shows that electron-like (purple) pockets and hole-like (yellow) pockets indeed coexist at the Fermi level, which is consistent with our Hall measurements shown in Fig.~\ref{Fig2}\textbf{a}. (2)~Our calculation shows that the electron-like pocket is indeed the Weyl cones (Fig.~\ref{Fig2}\textbf{g}). This agrees with our Landau fan diagram analysis (Fig.~\ref{Fig2}\textbf{d}) which suggests a linear band dispersion with a nontrivial Berry's phase. (3)~The calculated carrier density of the electron pockets is $5.07\times10^{17}$~cm$^{-3}$, which also agrees well with our experimentally measured value in Fig.~\ref{Fig2}\textbf{d}. (4)~Finally, the anisotropy of the Fermi surface area is found to be 4.9, which is also in line with the experimentally determined value of 5. All these agreements, taken collectively, provide compelling evidence that the electron-like bands that dominate the SdH oscillations are indeed the Weyl cones W1, and that the chemical potential is $\sim11.5$~meV above the energy of the Weyl nodes W1. We note that according to our band calculations, the energy position of the Weyl nodes W2 is $\sim13$~meV higher than that of W1. Therefore, based on our systematic measurements and our careful comparison with calculations, we obtain a band diagram presented in Fig.~\ref{Fig2}\textbf{f}. The chemical potential lies $\sim11.5$~meV above W1 but it is extremely close to W2. This means that although the contribution of Weyl cones W2 to the SdH oscillation (Fig.~\ref{Fig2}\textbf{a}, $\boldsymbol{H}\perp\boldsymbol{i}$, where $\boldsymbol{i}$ is the direction of the current) is not significant due to its extremely small size of Fermi surface, the Weyl cones W2 will play the most essential role in the chiral anomaly, the positive magnetoconductance at $\boldsymbol{H}\parallel\boldsymbol{i}$, because the chemical potential is extremely close to the Weyl nodes of W2.

Keeping the experimentally determined band diagram Fig.~\ref{Fig2}\textbf{f} in mind, we now turn to the measurements of the chiral anomaly. We first discuss how the chiral anomaly manifests itself in a magnetoconductance measurement in the presence of parallel electrical and magnetic fields ($\boldsymbol{H}\parallel\boldsymbol{i}$). In the semiclassical regime, in which our measurements are performed, the Weyl nodes act as sources of Berry flux in the BZ. This Berry flux alters the semiclassical equation of motion for the momentum $\boldsymbol{k}$ of an electronic wave-packet by an additional term $\propto(\boldsymbol{E}\cdot\boldsymbol{B})\boldsymbol{\Omega}_{\boldsymbol{k}}$~\cite{Duval06,Son13}, where $\boldsymbol{E}$ and $\boldsymbol{B}$ are the applied electric and magnetic fields and $\boldsymbol{\Omega}_{\boldsymbol{k}}$ is the momentum-dependent Berry field strength. Close to the Weyl node, which we assume to be isotropic for simplicity, $\boldsymbol{\Omega}_{\boldsymbol{k}}$ has the monopole-configuration $\boldsymbol{\Omega}_{\boldsymbol{k}}=\pm\boldsymbol{k}v_{\mathrm{F}}^3/(2|E_{\mathrm{F}}|^3)\propto E_{\mathrm{F}}^{-2}$, where the sign $\pm$ is the chirality of the Weyl node. Since the Weyl nodes are separated in momentum space, the scattering of electrons between them is suppressed and the extra contribution to the equations of motion alters the transport properties of the Weyl semimetal at nonzero applied $\boldsymbol{E}$ and $\boldsymbol{B}$ field as long as $\boldsymbol{E}\cdot\boldsymbol{B}\neq0$. For small magnetic fields parallel to $\boldsymbol{E}$, the contribution to the conductivity from an isotropic Weyl node is at low temperatures given by~\cite{Son13, Burkov14}
\begin{equation}
\sigma^{\mathrm{chiral}}\approx\sigma_0 \frac{2e^2 {\Omega_{\mathrm{F}}^2}}{3\hbar^2} B^2,
\end{equation}
where $\sigma_0$ is the conductivity originating from inter-node scattering and $\Omega_{\mathrm{F}}$ is the Berry field strength at the Fermi energy.
Because the Fermi energy of the Weyl nodes W1 is about 7 times larger than that of W2, the contribution of W1 to $\sigma^{\mathrm{chiral}}$ is suppressed by a factor of $\sim50$ as compared to the contribution from W2. We can thus assume that the chiral anomaly in TaAs is dominated by the Weyl node W2.
On the basis of these considerations, we expect $\sigma^{\mathrm{chiral}}$ to show the following behavior: 1) It has a weak temperature dependence for temperatures small compared to the $|E_{\mathrm{F}}|$ and is reduced for temperatures much larger than $|E_{\mathrm{F}}|$. 2) It is largest for $\boldsymbol{E}\parallel\boldsymbol{B}$ and zero for $\boldsymbol{E}\perp\boldsymbol{B}$. 3) It scales $\propto E_{\mathrm{F}}^{-2}$ for sufficiently small $E_{\mathrm{F}}$ \cite{Son13}. We will now show that our data is consistent with each of these points.

In the following, we present the magnetoconductivity instead of the magnetoresistance for two reasons: i) The conductivity is a sum of contributions from the chiral anomaly an conventional transport, which makes it easier to disentangle the two contributions. ii) The resistivity at zero field is dominated by the conventional contribution and it is therefore unnatural to normalize the contribution from the chiral anomaly that we want to single out by this nonuniversal value.

The magnetoconductivity shown in Fig.~\ref{Fig3} was measured in a configuration where the applied electric and magnetic fields are parallel. Figure~\ref{Fig4} shows the change in the magnetoconductivity as the angle between the applied electric and magnetic fields is varied. The measurements were performed with two samples, S1 and S2, that differ in the location of their Fermi  energy. Relative to the Weyl node W2, S1 has $E_{\mathrm{F},1}=-1.5\ \mathrm{meV}$ and S2 has $E_{\mathrm{F},2}=+3.0\ \mathrm{meV}$ (see Fig.~\ref{Fig3}\textbf{d}). For both samples, the magnetoconductivity in Fig.~\ref{Fig3}\textbf{c} shows two peaks as a function of magnetic field strength, one at zero and one at finite magnetic field. We identify these as a classical magnetoconductivity of conventional carriers and the chiral anomaly of the Weyl points, respectively.
To support this, we fitted the experimentally obtained conductivity with the functional form proposed in~\cite{Jho13}
\begin{equation}
\sigma(H)
=\left(\sigma_0+a\sqrt{H}\right)\left(1+C_WH^2\right)+\frac{1}{\rho+A\,H^2}+\frac{1}{\rho'+A'\,H^2},
\label{eq: fitfunction}
\end{equation}
where the first term is associated with the Weyl nodes and includes a weak anti-localization correction with coefficient $a<0$, the coefficient $C_W$ is due to the chiral anomaly, and the last two terms with parameters $A,A',\rho,\rho'>0$ are conventional contributions to the magnetoconductivity that we attribute to the other Fermi pockets. Using Eq.~\eqref{eq: fitfunction}, we obtain excellent fits to the data at all temperatures and angles, as presented in Fig.~\ref{Fig3}\textbf{a} and Fig.~\ref{Fig4}\textbf{a}, respectively. (We found that no satisfactory fit can be obtained, if only one of the conventional contributions is included.)

As a function of temperature, the chiral anomaly contribution $\sigma_0C_W$, shown in Fig.~\ref{Fig3}\textbf{b} for S1, is indeed nearly constant for small temperatures. It rapidly decreases for temperatures around $20\ \mathrm{K}$, which is comparable to the Fermi energy $|E_{\mathrm{F},1}|=1.5\,\mathrm{meV}$ of the sample. This confirms point 1) from the above. As a function of the angle between electric and magnetic field, the chiral anomaly contribution $\sigma_0C_W$, shown in Fig.~\ref{Fig4}\textbf{b} for S2, is found to be sharply peaked for $\boldsymbol{E}\parallel\boldsymbol{B}$ and falls off to nearly zero for relative angles as small as $\theta\sim5^\circ$. This is in qualitative agreement with point 2) from the above. Finally, with the two samples studied, we have observed the chiral anomaly contribution to the conductivity for two different Fermi energies, $E_{\mathrm{F},1}=-1.5\ \mathrm{meV}$ and $E_{\mathrm{F},2}=+3.0\ \mathrm{meV}$. We can use the two chiral anomaly contributions to the magnetoconductivity $\sigma_{0,1}C_{W,1}$ and $\sigma_{0,2}C_{W,2}$ obtained for S1 and S2 to test the scaling form $\sigma^{\mathrm{chiral}}\propto E_{\mathrm{F}}^{-2}$ quantitatively. Indeed, $\sigma_{0,1}C_{W,1}/(\sigma_{0,2}C_{W,2})=5.0$ (each taken at $T=2\,\mathrm{K}$) compares well with $E_{\mathrm{F},2}^2/E_{\mathrm{F},1}^2=4.0$, thus confirming point 3) from the above.

These systematic experimental results corroborated by our ARPES results and our theoretical calculations, suggest the existence of Weyl fermion related chiral anomaly in TaAs analogous to anomalies in quantum field theory. Our results not only provide the first transport signature for the Weyl fermions in TaAs, but also paves the way for realizing new device platforms for the next generation quantum technology frontier.

\bigskip
\bigskip
\textbf{Acknowledgements}
MZH thanks I. Klebanov and A. Polyakov for theoretical discussions and LBNL staff for experimental support. SJ thanks J. Xiong and F. Wang for valuable discussions and C.L. Zhang and Z.J. Yuan thank Yuan Li and Ji Feng for using instruments in their groups. S. J. is supported by National Basic Research Program of China (Grant Nos. 2013CB921901 and 2014CB239302). H.L. acknowledges the Singapore National Research Foundation for the support under NRF Award No. NRF-NRFF2013-03. The work at Princeton and Princeton-led synchrotron-based measurements is supported by U.S. DOE DE-FG-02-05ER46200 and by Gordon and Betty Moore Foundation grant for sample characterization.

\newpage
\addtocounter{figure}{-1}
\begin{figure*}[t!]
\caption{and blue lines represent the zeroth Landau level with + and - chiralities. \textbf{d,} ARPES Fermi surface map of the (001) cleaving plane of TaAs, clearly resolving Fermi arcs near the $\bar{X}$ point, $\bar{Y}$ point, midpoint of the $\bar{X}$ point and $\bar{\Gamma}$ point, and midpoint of the $\bar{Y}$ point and $\bar{\Gamma}$ point of the surface Brillouin zone. These momentum locations are indicated in \textbf{e}. \textbf{e,} An \textit{ab initio} band structure calculation of the surface states on the (001) surface of TaAs, in agreement with our experimental ARPES data. \textbf{f,} Distribution of the Weyl nodes throughout the first bulk Brillouin zone (BZ), where a total of 24 Weyl nodes are present and the sign of their chiral charges are color coded in black and white. 8 of these Weyl nodes are located on the $k_z$ = $\pi$ plane and have the same energy (W1), whereas the other 16 Weyl nodes are away from the $k_z$ = $\pi$ plane (W2). \textbf{g,} Temperature dependence of resistivity at different magnetic fields perpendicular to the current. \textbf{h,} Magneto-resistance (MR) at various temperatures. \textbf{i,} Schematics of the experimental setup showing the directions of the magnetic field and electric current. \textbf{j,} MR of the sample in a pulse magnetic field as high as 56 T. The red line is the fitting to the experimental result. All transport data are obtained from sample S1. [Chenglong Zhang, Su-Yang Xu, Ilya Belopolski, Zhujun Yuan \textit{et.al.,} (2015)]}
\end{figure*}

\clearpage

\begin{figure*}
\centering
\includegraphics[width=17cm]{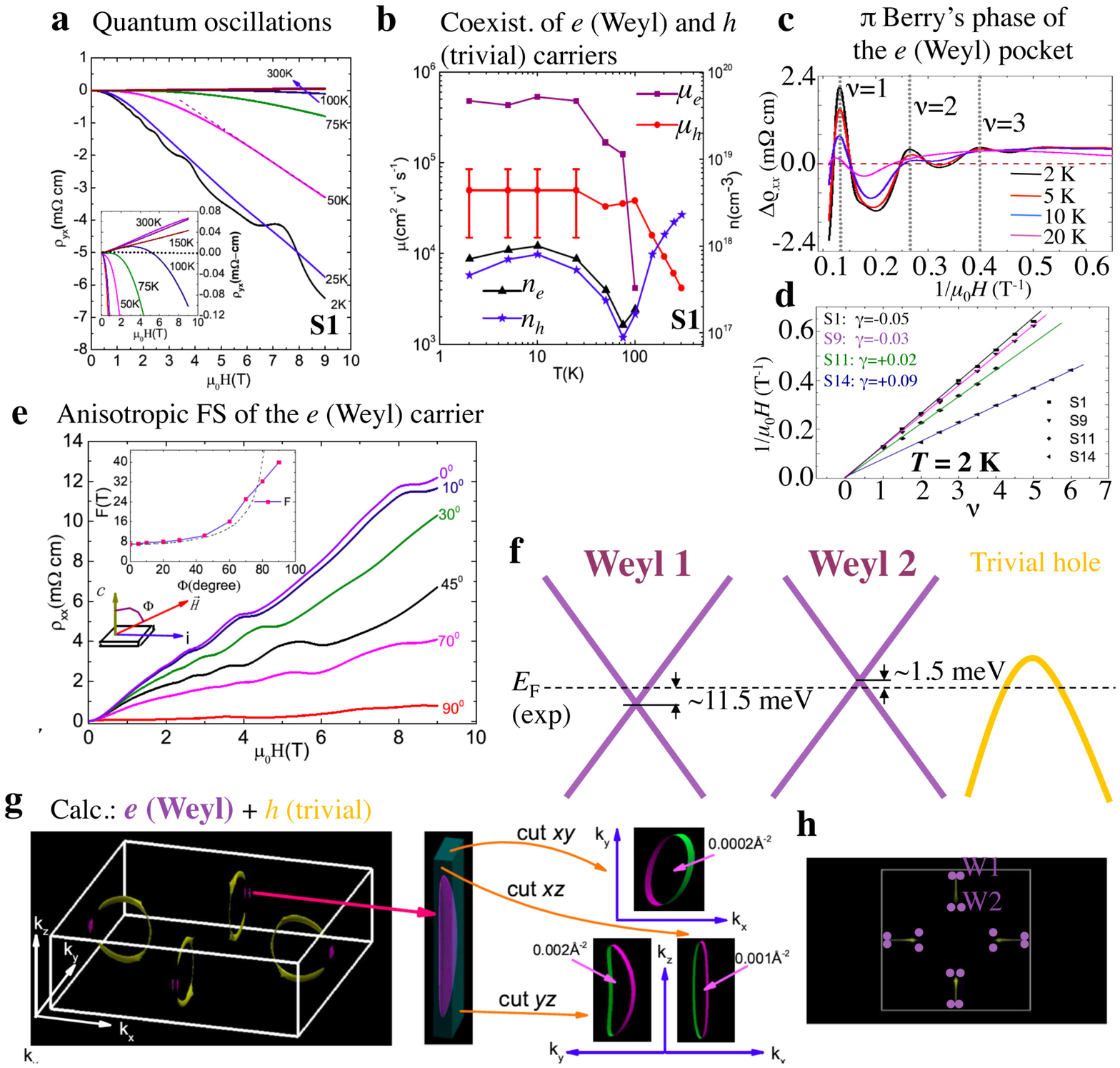}
\caption{\label{Fig2}\textbf{Probing the Weyl bands via quantum oscillation measurements.} \textbf{a,} Hall resistivity versus magnetic field in the temperature range from 2 to 300 K. Strong SdH oscillations were observed at 2 K. Inset: the high temperature Hall resistivity. \textbf{b,} Mobilities and carrier concentrations of the electrons and holes, clearly showing the coexistence of Weyl electrons and trivial holes in our samples. No information of the electrons can be obtained above 100 K in this measurement. \textbf{c,} The oscillatory parts of $\sigma_{xx}$ at various temperatures, showing the $\pi$ Berry's phase of the Weyl electron pocket. \textbf{d,} The SdH fan diagram for four different samples. All of the four intercepts are located around zero, suggesting the $\pi$ Berry’s phase of the Weyl electron pocket \textbf{e,} Magnetic field dependence of resistivity at representative $\Phi$ angles}
\end{figure*}
\addtocounter{figure}{-1}
\begin{figure*}[t!]
\caption{between $0^{\circ}$ - $90^{\circ}$ at 2K for S1, after heating the sample. The MR decreases rapidly when the magnetic field is tilted from \textit{c} to the direction of the current \textit{i}. Inset: the frequency F versus $\Phi$. The dashed curve is (1/$\cos\Phi$)$\cdot$F$_{0}$. \textbf{f,} Schematics of the experimentally determined band diagram indicating the positions of the two different types of Weyl nodes, W1 and W2, as well as the trivial hole-like band, relative to the Fermi level. \textbf{g,} (left) Shapes and locations of electron and hole pockets from first-principles calculations. (center) The magnified electron pocket near the Weyl point 1 (W1). (right) Three extremal cross-section areas along the three crystallographic axes. \textbf{h,} Schematics of the distribution of the W1 and W2 Weyl nodes in first Brillouin zone. [Chenglong Zhang, Su-Yang Xu, Ilya Belopolski, Zhujun Yuan \textit{et.al.,} (2015)]}
\end{figure*}

\begin{figure*}
\centering
\includegraphics[width=17cm]{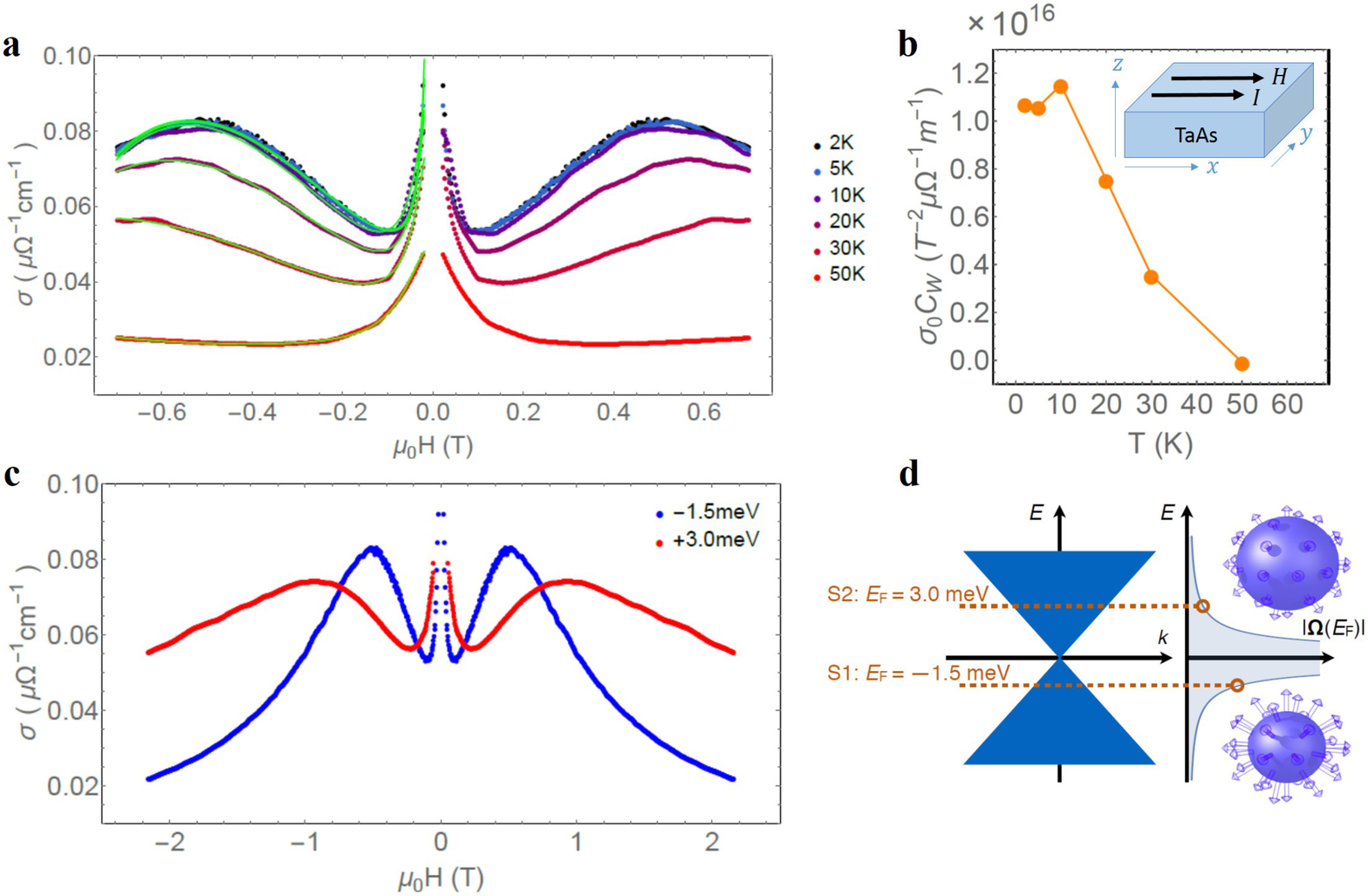}
\caption{\label{Fig3}\textbf{Observation of the chiral anomaly in the magnetoconductivity.}
\textbf{a,} Magnetoconductivity for sample S1 at different temperatures and fits to the data using $\sigma(H)$ defined in Eq.~\eqref{eq: fitfunction}. The peaks at zero and finite magnetic fields are interpreted as the classical and the chiral anomaly contribution to the magnetoconductivity, respectively. \textbf{b,} Temperature-dependence of the chiral anomaly contribution to the magnetoconductivity as obtained from the fits in panel (\textbf{a}). A saturation of the chiral anomaly contribution to the magnetoconductance occurs below the temperature that corresponds to the Fermi energy, temperatures below the Fermi energy is observed, matching the theoretical expectation. \textbf{c,} Magnetoconductivity at $T=2$~K for two different samples S1 and S2. \textbf{d,} Locations of the chemical potentials relative to the Weyl node W2 in samples S1 and S2 and the schematic dependence of the magnitude of the Berry curvature $|\boldsymbol{\Omega}_{\boldsymbol{k}}|$ at the Fermi energy on the distance from the Weyl node. The Berry curvature is in a hedgehog-like monopole configuration on the Fermi surface. [Chenglong Zhang, Su-Yang Xu, Ilya Belopolski, Zhujun Yuan \textit{et.al.,} (2015)]}
\end{figure*}

\begin{figure*}
\centering
\includegraphics[width=17cm]{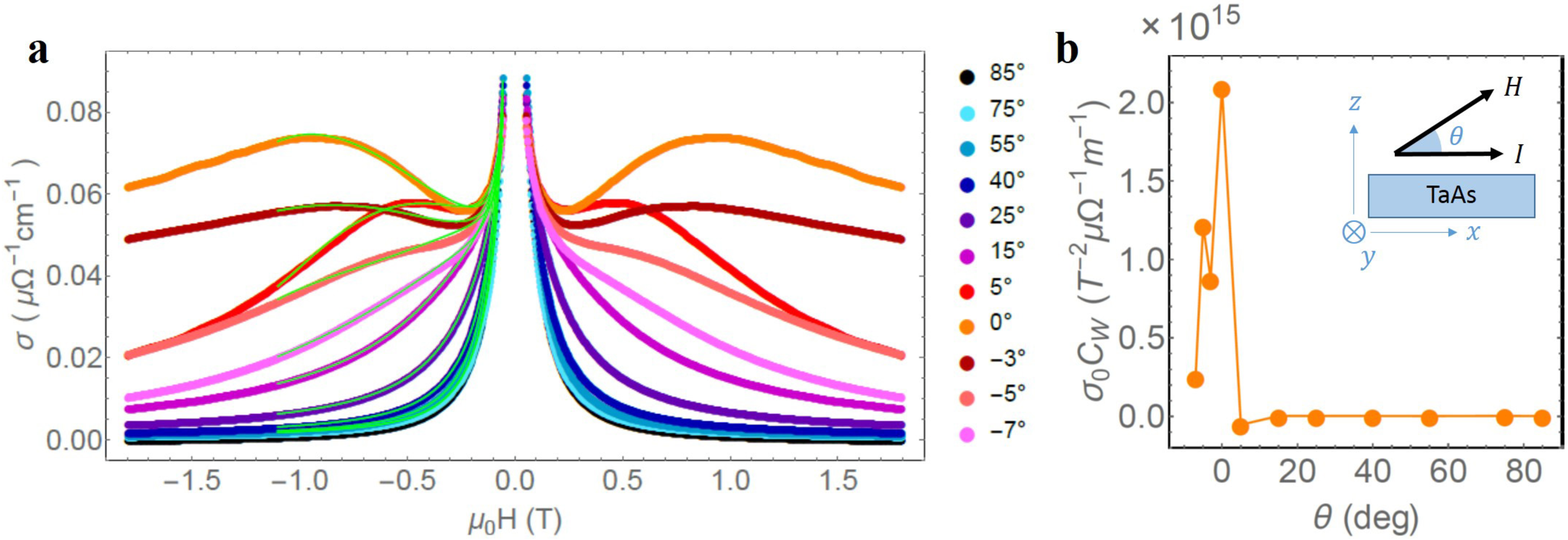}
\caption{\label{Fig4}\textbf{Dependence of the chiral anomaly on the angle between applied electric and magnetic field.} \textbf{a,} Magnetoconductivity for sample S2 at different angles and $T=2$~K as well as fits to the data using $\sigma(H)$ defined in Eq.~\eqref{eq: fitfunction}. \textbf{b,} Angle-dependence of the chiral anomaly contribution to the magnetoconductivity as obtained from the fits in panel (\textbf{a}). In line with the theoretical expectation, the chiral anomaly is largest when electric and magnetic fields are parallel. As the field is tilted, the chiral anomaly contribution rapidly decreases to zero. [Chenglong Zhang, Su-Yang Xu, Ilya Belopolski, Zhujun Yuan \textit{et.al.,} (2015)]}
\end{figure*}

\end{document}